\documentstyle[stwol]{article}

\input{psfig}

\evensidemargin\oddsidemargin

\def\be{\begin{equation}}
\def\ee{\end{equation}}
\def\bea{\begin{eqnarray}}
\def\eea{\end{eqnarray}}

\def\er#1#2{\relax\ifmmode{}^{+#1}_{-#2}\else$^{+#1}_{-#2}$\fi}
\def\erparen#1#2{\relax\ifmmode{}(^{#1}_{#2})\else$(^{#1}_{#2})$\fi}
\def\btorho{\bar B^0{\to}\rho^+ l^- \bar\nu_l}
\def\btopi{\bar B^0{\to}\pi^+ l^- \bar\nu_l}
\def\btokstargamma{\bar B{\to} K^* \gamma}
\def\qsqmax{q^2_{\rm max}}
\def\vub{|V_{ub}|}
\def\w{\omega}

\def\alphas{\alpha_{\rm s}}
\def\alphaslatt{\alpha_{\rm s}^{\rm latt}}
\def\alphasmsbar{\alpha_{\rm\overline{MS}}}
\def\alphaP{\alpha_{\rm P}}
\def\bkhat{\hat B_K}
\def\bktwo{B_K^{\rm NDR}(2\gev)}
\def\mmsbar{m^{\rm\overline{MS}}}
\def\bra#1{\left\langle #1 \right|}
\def\ket#1{\left| #1 \right\rangle}
\def\gev{\,{\rm Ge\kern-0.1em V}}
\def\mev{\,{\rm Me\kern-0.1em V}}
\def\ts{\vrule height3ex depth0pt width0pt}

\def~{\ifmmode\phantom{0}\else\penalty10000\ \fi}
\newdimen\unit
\def\point#1 #2 #3{\vbox to0pt{\kern-#2\unit
  \hbox{\kern#1\unit$#3$}\vss}
 \nointerlineskip}
\begin{document}

\title{\hfill
DEVELOPMENT IN LATTICE QCD\hfill
\raisebox{5ex}{\makebox[0pt][r]{\rm SHEP--96--33}}}

\author{JONATHAN M FLYNN}

\address{Department of Physics, University of Southampton, Highfield,
Southampton, SO17 1BJ, UK}

\twocolumn[\maketitle\abstracts{After a brief discussion of the
promise and limitations of the lattice technique, I review lattice QCD
results for several quantities of phenomenological interest. These
are: matrix elements for heavy-to-light meson decays, leptonic decay
constants $f_B$ and $f_D$, the parameters $B_B$ and $B_K$ for neutral
$B$ and $K$ meson mixing respectively, the strong coupling constant,
light quark masses and the lightest scalar glueball mass.}]

\section{Introduction}

In this review I concentrate on lattice QCD results for a selection of
quantities directly relevant for phenomenology, some of which were not
otherwise reported at this conference. I apologise to presenters of
lattice studies which I do not have space to report on here and refer
the interested reader directly to the parallel session
reports.\renewcommand{\thefootnote}{\relax}\footnote{Plenary talk at
ICHEP 96, 28th International Conference on High Energy Physics,
Warsaw, Poland, 25--31 July 1996}\addtocounter{footnote}{-1}
\renewcommand{\thefootnote}{\alph{footnote}}

Lattice QCD is an important tool for the non-perturbative evaluation
of strong interaction effects, but a wary consumer should keep in mind
the main sources of error, which I will briefly mention below, when
using lattice results. One beauty of the lattice approach is that
these errors can be systematically investigated and reduced.

The presentation begins with form factors for $\btopi$, $\btorho$ and
$\btokstargamma$. The semileptonic decays are important for
determining $\vub$; the radiative decay for $V_{ts}$ and as a window
on new physics. Next follow results for the $B$ meson decay constant
$f_B$ and mixing parameter $B_B$, crucial for constraining the
Standard Model unitarity triangle. I then turn to the kaon mixing
parameter $B_K$, strong coupling $\alphas$ and the light quark masses,
concluding with results for the lightest scalar glueball.

\section{Lattice Calculations}

The standard lattice approach to QCD uses a discretised finite volume
region of Euclidean space-time on which the quantum field theory path
integral becomes a well-defined multi-dimensional integral evaluated
by Monte Carlo methods.

Current simulations use lattice spacings $a$ in the range
$0.05$--$0.2\,\rm fm$, sufficient to cover a hadron of size $0.5\,\rm
fm$ or greater by a few lattice points. These $a$'s correspond to
energy scales of $1$--$4\gev$ putting the lattice ultraviolet cutoff
above the scale of low energy QCD dynamics. Continuum results should
be obtained in the limit $a{\to}0$: this continuum extrapolation is
now feasible for many quantities.  Lattices should be large enough
that hadrons will comfortably fit on them. Current spatial sizes are
of order $2\,\rm fm$ or so, though some practitioners advocate at
least $2.5\,\rm fm$.\cite{gottlieb:lat96}

Because the QCD action is quadratic in the quark fields, matrix
elements of quarks can be evaluated using quark propagators in the
gluon background together with the gluon-field-dependent determinant
of the fermion operator.  The determinant is extremely demanding to
calculate so it is often set to its average value in the gauge field
background---the quenched approximation---which corresponds to
neglecting internal quark loops. In practice, much of the effect of
internal loops is to change the running of the coupling constant and
this can be compensated by changing the value of $\beta$, the bare
coupling which is input. Quenching is one of the systematic effects
causing disagreement between lattice spacings determined from
different physical quantities.

Calculating quark propagators means inverting the fermion
operator. This is very slow for realistic light quarks, so a range of
masses around the strange mass is simulated and then a `chiral
extrapolation' made to realistic values. For heavy quark masses $m$,
the inversion is fast but for large $ma$ discretisation errors are
large. Hence, $B$ physics results using relativistic fermion actions
typically involve an extrapolation from results at masses close to the
charm scale. An alternative is to use static or nonrelativistic
(NRQCD) lattice actions. In the static case the quark mass is treated
analytically, outside the simulation, and results are obtained in a
systematic expansion around the infinite mass limit. NRQCD actions
work well for quarks around the $b$ mass and above, but begin to fail
as one approaches the $c$ mass. Actions are also used which
interpolate between the static and relativistic
extremes.\cite{fermilab-action}

Lattice calculations provide matrix elements of bare operators defined
with the lattice regularisation, but results are needed for matrix
elements in a continuum scheme, like $\overline{\rm MS}$. A continuum
operator typically matches onto a set of lattice operators,
\[
{\cal O}_i^{\rm cont}(\mu) = Z_{ij}(\mu,a) {\cal O}_j^{\rm latt}(a),
\]
where the renormalisation constants $Z_{ij}$ can be calculated
perturbatively. Lattice perturbation theory using the bare coupling
$g_0^2 = 6/\beta$ is notoriously badly-behaved, however.~\cite{lm}
Abandoning $g_0^2$ in favour of a continuum-like definition leads to
better behaviour, but nonperturbative methods are being developed and
used (some are mentioned below).\cite{rossi:lat96} This looks to be
the way of the future.

Lattice gluon actions have discretisation errors beginning at
$O(a^2)$. The simplest, Wilson, fermion action, a latticised
relativistic Dirac action, has errors beginning at $O(a)$. One can
simply accept the $O(a)$ errors and let the $a{\to}0$ extrapolation
remove them. However, removing or reducing the $O(a)$ errors allows
simulations at coarser lattice spacings and makes the continuum
extrapolation less severe. This `improvement' of lattice actions is
currently a very active subject. The
Sheikholeslami-Wohlert~\cite{SWfermions} (SW or `clover') action
removes $O(a)$ errors at tree level, so the first corrections are
$O(\alphas a)$. The SW action contains one extra term compared to the
Wilson action. By tuning the coefficient of this term one can reduce
or even remove all $O(a)$ errors in physical quantities (one has to
improve operators used in matrix element calculations also). Partial
removal using `tadpole-improvement'~\cite{lm} is being widely applied,
but an ambitious program by the ALPHA
collaboration~\cite{alpha:lat96,pa15-010} aims for complete removal.

\section{Semileptonic and Radiative Heavy-to-light Decays}

Lattice form factor calculations are crucial here: the overall
normalisation at the zero recoil point $\w = v{\cdot}v' = 1$ is not
fixed by heavy quark symmetry as it is for a heavy-to-heavy
transition. Here $v$ and $v'$ are the four-velocities of the $B$ meson
of mass $M$ and the meson of mass $m$ it decays into,
respectively. The squared momentum transfer to the leptons or
photon is $q^2 = M^2 + m^2 - 2Mm\w$.

\begin{figure}
\hbox to\hsize{\hfill\vbox{\offinterlineskip
\psfig{figure=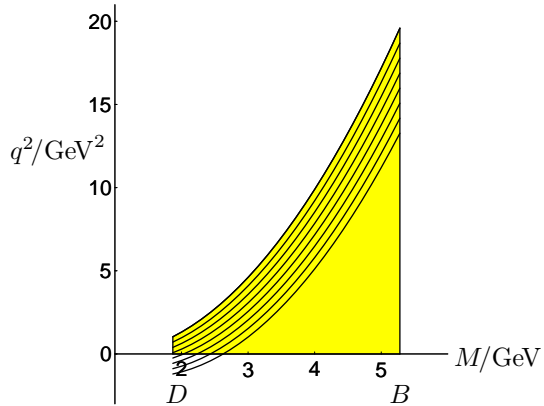,width=0.7\hsize}
\unit=0.7\hsize
\point -0.03 0.66 {q^2\!/\!\gev^2}
\point 1 0.17 {M\!/\!\gev}
\point 0.33 0.08 D
\point 0.85 0.08 B
}\hfill}
\caption[]{$q^2$ range for heavy-to-light decays as a function of the
decaying heavy meson mass. Lines of constant $\w$ are shown. The light
final state mass is taken to be $850\mev$, typical of lattice
pseudoscalar or vector meson masses before chiral extrapolation.}
\label{fig:heavy-to-light-scaling}
\end{figure}
Relativistic quark calculations use heavy quark masses around
the charm mass. The initial heavy meson is given $0$ or $1$ lattice
units of three-momentum, while the light final meson can generally be
given up to two lattice units of spatial momentum, allowing $q^2$ to
be varied from $\qsqmax = (M{-}m)^2$ (where $\w=1$) down to $q^2 <0$
at the $D$ scale.  Heavy quark symmetry determines the $M$ dependence
of the form factors at fixed $\w$, enabling an extrapolation to the
$B$ scale.  Fig.~\ref{fig:heavy-to-light-scaling} shows that scaling
in $M$ at fixed $\w$ sweeps all the measured points to a region near
$\qsqmax$ for $B$ decays. The problem is then to extrapolate back down
to $q^2=0$. This is particularly acute for the radiative decay
$\btokstargamma$ where only the form factors at $q^2=0$
contribute. Even if a static, non-relativistic or other modified
action is used for the heavy quark, the restriction on usable
three-momenta on current lattices, caused by momentum-dependent errors
and increasing statistical uncertainty, ensures that form factor values are
obtained only near $\qsqmax$.\cite{simone:lat95}

Some assistance is provided by ensuring that any model $q^2$
dependences respect heavy quark symmetry and constraints relating form
factors at $q^2=0$. For example, $f^+$ and $f^0$ for $B\to\pi$ are
related at $q^2=0$ and consistency can be achieved by fitting $f^+$ to
a dipole [pole] form and $f^0$ to a pole [constant] form. Heavy quark
symmetry and light flavour $SU(3)$ symmetry relate the $\btorho$ and
$\btokstargamma$ form factors. Models can further relate these to form
factors for $\btopi$. An overall fit might then be used. However, it
is clearly desirable to avoid models entirely.
\begin{figure*}
\hbox to\hsize{\hss\vbox{\offinterlineskip
\psfig{figure=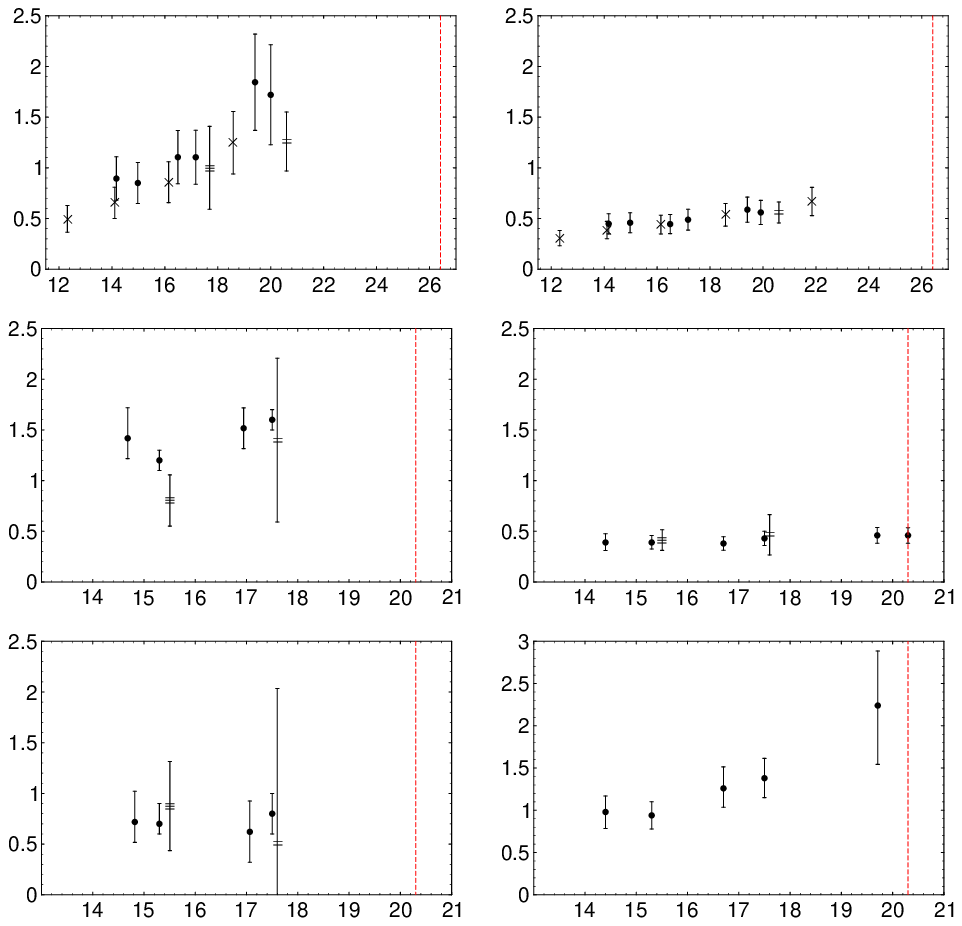,width=0.85\hsize}
\unit=0.85\hsize
\point 0.87 0.03 {q^2/\!\gev^2}
\point 0.055 0.965 {f^+}
\point 0.055 0.64 V
\point 0.055 0.31 {A_2}
\point 0.57 0.965 {f^0}
\point 0.57 0.64 {A_1}
\point 0.57 0.31 {A_0}
}\hss}
\caption[]{Lattice measurements of the form factors for $\btopi$ and
$\btorho$. Points are from ELC~\cite{elc:hl-semilept} ($\equiv$),
APE~\cite{ape:hl-semilept,vl} ($=$), UKQCD~\cite{ukqcd:hlff}
($\bullet$) and FNAL~\cite{simone:lat95,fnal:fBfD-lat95} ($\times$,
preliminary). Systematic errors have been
added~\cite{lpl:btopi-bounds} to the UKQCD and FNAL points. The
vertical dashed lines show $\qsqmax$.}
\label{fig:btopirho-form-factors}
\end{figure*}

\subsection{Semileptonic $B \to \pi$}

Lattice results~\cite{elc:hl-semilept,ape:hl-semilept,vl,ukqcd:hlff}
for the $\btopi$ form factors, $f^+$ and $f^0$, are plotted in
Fig.~\ref{fig:btopirho-form-factors} (further
results~\cite{wup:sl-lat95,wup:hl-weak-decays} at $q^2=0$ only are not
displayed).

\begin{figure}
\hbox to\hsize{\hss\vbox{\offinterlineskip
\psfig{file=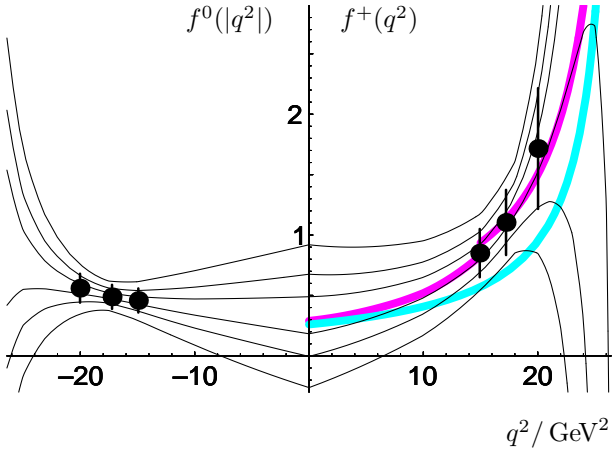,width=\hsize}
\unit=\hsize
\point 0.55 0.74 {f^+(q^2)}
\point 0.3 0.74 {f^0(|q^2|)}
\point 0.82 0.07 {q^2/\gev^2}
}\hss}
\caption[]{Bounds on $f^+$ and $f^0$ for $\btopi$ from dispersive
constraints.\cite{lpl:btopi-bounds} The data points are from
UKQCD,\cite{ukqcd:hlff} with added systematic errors. The pairs of
fine curves are, outermost to innermost, 95\%, 70\% and 30\% bounds.
The upper and lower shaded curves are
light-cone~\protect\cite{bbkr:B-Bstar-pi-couplings} and
three-point~\protect\cite{PaB93} sum rule results respectively.}
\label{fig:btopi-bounds}
\end{figure}

For massless leptons, the decay rate is determined by $f^+$
alone. However, the constraint $f^+(0) = f^0(0)$ (with suitable
conventions) makes lattice measurements of both form factors useful.
One procedure uses dispersive constraints, obtained by combining
dispersion relations with unitarity, analyticity, crossing symmetry
and perturbative QCD, to obtain model independent bounds.
Lellouch~\cite{lpl:btopi-bounds,lpl:ichep96} has shown how to
incorporate the $f^+(0)=f^0(0)$ condition together with imperfectly
known values of the form factors, typical of lattice results with
errors, to obtain families of bounds with varying confidence levels. A
set of such bounds is shown in Fig.~\ref{fig:btopi-bounds} together
with the UKQCD~\cite{ukqcd:hlff} lattice measurements used to obtain
them. In the figure, $f^0$ and $f^+$ are plotted back-to-back, showing
the effect of imposing the constraint at $q^2=0$.

In Table~\ref{tab:btopi-results} the bounds have been used to give
ranges of values for the decay rate $\Gamma(\btopi)$ in units of
$\vub^2\,{\rm ps}^{-1}$ together with values for the form factor $f^+$
at $q^2=0$. When combined with the experimental result for the decay
rate, one can extract $\vub$ with about 35\% theoretical
error. Although not very precise, this result relies only on lattice
calculations of matrix elements and heavy quark symmetry, together
with perturbative QCD and analyticity properties in applying the
dispersive constraints. There is no model dependence. Improved lattice
results can be used as input for the bounds once they become
available.

Also shown in Table~\ref{tab:btopi-results}, for comparison, are
values obtained from lattice calculations where assumed $q^2$
dependences have been imposed. For ELC~\cite{elc:hl-semilept} and
APE,\cite{ape:hl-semilept} one value of $f^+$ has been used, at the
given value of $q^2$, fitted to a single pole form with pole mass
$m_{\rm p}$. The UKQCD result~\cite{ukqcd:hlff} is obtained from a
combined dipole/pole fit to all measured $f^+$/$f^0$ points. Note that
the UKQCD points have statistical errors only and have not been
chirally extrapolated---they correspond to a pion mass of around
$800\mev$ (a similar caveat applies for the
FNAL~\cite{simone:lat95,fnal:fBfD-lat95} results for $f^+$ and
$f^0$). The results given~\cite{drb} have used these values as though
they applied to the physical pion. In obtaining bounds based on these
points, Lellouch~\cite{lpl:btopi-bounds} added a conservatively
estimated systematic error including terms to account for the chiral
extrapolation (this error has been added to the UKQCD and FNAL points
plotted in Fig.~\ref{fig:btopirho-form-factors}).
\begin{table}
\caption[]{$\btopi$ results from dispersive constraints applied to
lattice data,\cite{lpl:btopi-bounds} together with results
obtained using ans\"atze for the form factor $f^+$. The decay rates
are values for $\Gamma(\btopi)/\vub^2 {\rm ps}^{-1}$}
\label{tab:btopi-results}
%\vspace{0.4cm}
\begin{center}
\begin{tabular}{llll} \hline
\ts  & Rate & $f^+(0)$ & \\[1ex]
\hline
\ts Dispersive  & $2.4$--$28$ & $-0.26$--$0.92$ & 95\% CL\\
Constraint~\cite{lpl:btopi-bounds}
 & $3.6$--$17$ & $0.00$--$0.68$ & 70\% CL\\
 & $4.8$--$10$ & $0.18$--$0.49$ & 30\% CL\\[1ex]
\hline
\ts ELC~\cite{elc:hl-semilept} & $9\pm 6$ & 0.10--0.49 & \\
\multicolumn{4}{l}{$q^2{\simeq} 18\gev^2$, pole fit,  
$m_{\rm p}{=}5.29(1)\gev$}\\[1ex]
\hline
\ts APE~\cite{ape:hl-semilept} & $8\pm 4$ & 0.23--0.43 & \\
\multicolumn{4}{l}{$q^2{\simeq} 20.4\gev^2$, pole fit,
$m_{\rm p}{=}5.32(1)\gev$}\\[1ex]
\hline
\ts UKQCD~\cite{ukqcd:hlff,drb} & $7 \pm 1$ & 0.21--0.27 & \\
\multicolumn{4}{l}{dipole/pole fit to $f^+$/$f^0$
}\\[1ex]
\hline
\end{tabular}
\end{center}
\end{table}

Chiral extrapolations are severe for the $B\to\pi$ matrix elements. As
the pion mass approaches its physical value the $B^*$-pole and the
beginning of the $B\pi$ continuum approach $\qsqmax$ from above and
the form factors may vary rapidly with the pion mass near
$\qsqmax$. This is not a problem for $B\to\rho$ so lattice
calculations of semileptonic $B\to\rho$ decay are currently most
reliable and I now turn to them.

\subsection{Semileptonic $B \to \rho$}

To avoid models for the $q^2$ dependence of the form factors, we use
the lattice results directly. The lattice can give the
differential decay rate $d\Gamma/dq^2$, or the partially integrated
rate, in a $q^2$ region near $\qsqmax$ up to the unknown factor
$\vub^2$. For example, UKQCD~\cite{ukqcd:btorho} parametrised the
differential decay rate near $\qsqmax$ by,
\begin{equation}
{d\Gamma\over dq^2} = {\rm const} |V_{ub}|^2 q^2 \lambda^{1/2} a^2
\big(1+b[q^2{-}\qsqmax]\big),\label{eq:dGdqsq}
\end{equation}
where $\lambda$ is the usual phase space factor and $a$ and $b$ are
constants. The constant $a$ plays the role of the Isgur-Wise function
evaluated at $\w=1$ for heavy-to-heavy transitions, but in this case
there is no symmetry to determine its value.

For massless leptons, the differential decay rate depends on the $V$,
$A_1$ and $A_2$ form factors, but $A_1$ is the dominant contribution
near $\qsqmax$ and is also the best measured on the lattice, as shown
in Fig.~\ref{fig:btopirho-form-factors}. UKQCD performed a fit to the
parametrisation in Eq.~(\ref{eq:dGdqsq}) to find,\cite{ukqcd:btorho}
\[
a = 4.6 \er{0.3}{0.4} \pm 0.6 \gev.
\]
Discounting experimental errors, this result will allow determination
of $\vub$ with a theoretical uncertainty of 10\% statistical and 12\%
systematic.  CLEO are beginning to extract the differential decay
distributions.\cite{lkg:ichep96}

The UKQCD results for $V$, $A_1$ and $A_2$ agree very well with a
light cone sum rule (LCSR) calculation of Ball and
Braun.\cite{ball:lcsr-moriond} More interestingly, LCSR calculations
predict that all the form factors for heavy-to-light decays have the
following heavy mass dependence at
$q^2=0$:~\cite{bbkr:B-Bstar-pi-couplings,cz:lcsr-orig}
\begin{equation}
f(0) = {1\over M^{3/2}} (a_0 + a_1/M + a_2/M^2 + \cdots).
\label{eq:lcsr-ff-zero}
\end{equation}
The leading $M$ dependence comprises $\sqrt{M}$ from the heavy state
normalisation together with the behaviour of the leading twist light
cone wavefunction. For $A_1(0)$ the LCSR result
is~\cite{ball:lcsr-moriond} $A_1(0) \simeq 0.26$. UKQCD fitted the
heavy mass dependence of $A_1(0)$ to the form in
Eq.~(\ref{eq:lcsr-ff-zero}) and found~\cite{ukqcd:btorho}
\[
A_1(0) = \cases{0.18\pm0.02&$a_0$ and $a_1 \neq 0$\cr
0.22\er{0.04}{0.03}&$a_0$, $a_1$ and $a_2 \neq 0$\cr}
\]
Pole fits for $A_1$ have leading $M^{-3/2}$ behaviour at $q^2=0$, so
we can also compare with other lattice results:
\[
A_1(0) = \cases{0.22\pm0.05&ELC \cite{elc:hl-semilept}\cr
0.24\pm0.12&APE \cite{ape:hl-semilept}\cr
0.27\er74&UKQCD \cite{ukqcd:btorho}\cr}
\]

\subsection{Rare radiative $B \to K^* \gamma$}

This decay was discussed in some detail by A.~Soni at Lattice
95~\cite{soni:lat95} so my comments will be brief.
Table~\ref{tab:btokstargamma} summarises available lattice results,
all from quenched simulations, for the matrix element
\begin{equation}\label{eq:bkstarME}
\langle K^*(k,\eta) | \overline{s} \sigma_{\mu\nu} q^\nu b_R
 | B(p) \rangle
\end{equation}
which is parameterised by three form factors, $T_i$, $i=1,2,3$. For
the decay rate, the related values $T_1(0)$ and $T_2(0)$ are
needed. Suitably defined, they are equal, so the table lists a single
value $T(0)$, together with the directly measured $T_2(\qsqmax)$. The
results are classified by the leading $M$ dependence of the form
factor at $q^2=0$ which is governed by the model used for the $q^2$
dependence. Dipole/pole forms for $T_1$/$T_2$ give $M^{-3/2}$
behaviour and pole/constant forms give $M^{-1/2}$. The results agree
when the same assumptions are made. All groups find that $T_2$ has
much less $q^2$ dependence than $T_1$, but the overall forms cannot be
decided, so a phenomenological prediction is elusive.
\begin{table}
\caption[]{Lattice results for $\btokstargamma$.}
\label{tab:btokstargamma}
\vspace{0.4cm}
\begin{center}
\begin{tabular}{llll}\hline
\ts & \multicolumn{2}{c}{$T(0)$} \\[0.5ex]
\cline{2-3}
\ts & $M^{-3/2}$ & $M^{-1/2}$ & $T_2(\qsqmax)$
\\[0.5ex]
\hline
\ts BHS~\cite{bhs:bsg} & 0.10(3) &         & 0.33(7) \\
LANL~\cite{lanl:wme-lat95}   & 0.09(1) & 0.24(1) & \\
%APE~\cite{ape:bsg-clover}    & 0.09(1)(1) & 0.23(2)(2) & 0.23(2)(2) \\
APE~\cite{ape:bsg-clover}    & 0.09(1) & 0.23(3) & 0.23(3) \\
UKQCD~\cite{ukqcd:hlff} & 0.15\erparen76 & 0.26\erparen21 & 0.27\erparen21 \\
BHS~\cite{bhs:bsg}   & & & 0.30(3)\\[0.5ex]
\hline
\end{tabular}
\end{center}
\end{table}

Additional long distance contributions may not be negligible so the
matrix element of Eq.~(\ref{eq:bkstarME}) may not give the true decay
rate.\cite{gp:longdist1,gp:longdist2,abs} Once the $q^2$ dependence of
the form factors is known, lattice calculations of the ratio $R_{K^*}
= \Gamma(\btokstargamma)/\Gamma(b\to s\gamma)$ can be compared to the
experimental result to test for long distance effects.

%Heavy quark symmetry, combined with light flavour $SU(3)$ symmetry,
%relates the form factors for $\btorho$ and
%$\btokstargamma$~\cite{iw:hqet,gmm} in the infinite heavy quark mass
%limit. On the lattice these relations can be tested using identical
%light quarks. UKQCD~\cite{ukqcd:btorho} showed that the ratios
%$V/2T_1$ and $A_1/2T_2$ both satisfied the heavy quark symmetry
%prediction of unity in the infinite mass limit. A combined fit of the
%pseudoscalar to vector form factors consistent with heavy quark
%symmetry could help resolve the ambiguity in the $q^2$ dependence of
%the $\btokstargamma$ form factors.

\section{Leptonic Decay Constants $f_B$ and $f_D$}

\begin{table*}[t]
\caption[]{Values and ratios of leptonic decay constants of
pseudoscalar mesons from recent continuum extrapolations by
MILC~\cite{milc:fb-lat96} and JLQCD~\cite{jlqcd:fb-lat96}. Statistical
and systematic errors have been combined in quadrature. The MILC
results include a systematic error for quenching. The bottom row
contains summary values based on global
data.\cite{jmf-hlwme:lat96,marti:heavy-flavours,marti:beauty96,marti:ichep96}}
\label{tab:pscalar-decay-consts}
\vspace{0.4cm}
\begin{center}
\begin{tabular}{lllllll}
\hline
\ts &
$f_B/\mev$ & $f_{B_s}/\mev$ & $f_{B_s}/f_B$ &
$f_D/\mev$ & $f_{D_s}/\mev$ & $f_{D_s}/f_D$ \\[1ex]
\hline
\ts MILC~\cite{milc:fb-lat96} & 166(33) & 181(41)
& 1.10(10) & 196(18) & 211(28) &
1.09(7) \\
JLQCD~\cite{jlqcd:fb-lat96} & 179\erparen{11}{33} &
197\erparen{~7}{36} &
 & 202\erparen{25}{14} & 216\erparen{23}{16}\\[0.5ex]
\hline
\ts Summary
 & 175(25) & 200(25) & 1.15(5) & 205(15) & 235(15) & 1.15(5)\\[0.5ex]
\hline
\end{tabular}
\end{center}
\end{table*}
The leptonic decay constant of a pseudoscalar meson $P$ is defined by
the axial current matrix element
\[
\bra0 A_\mu(0) \ket{P(p)} = i f_P p_\mu.
\]
On the lattice one calculates a dimensionless quantity $Z_L$, from
which $f_P$ is obtained via
\[
f_P \sqrt{M_p/2} = Z^{\rm ren} Z_L a^{-3/2}
\]
where $Z^{\rm ren}$ is the renormalisation constant required to match
to the continuum and $a$ is the lattice spacing.

Two collaborations have new values, shown in
table~\ref{tab:pscalar-decay-consts}, for $f_B$ and $f_D$ obtained
from continuum extrapolations of quenched results at several
lattice spacings. JLQCD~\cite{jlqcd:fb-lat96} study different
prescriptions for reducing the $O(ma)$ discretisation errors
associated with heavy quarks and aim to show that all results converge
in the continuum limit. Their error is statistical combined with an
error from the spread over prescriptions.

MILC~\cite{milc:fb-lat96} simulate a range of heavy quark masses plus
a static (infinite mass) quark, giving meson masses straddling $m_D$
and allowing an interpolation to $m_B$. Their (preliminary) results
give errors from: (i) statistics, (ii) systematics within the quenched
approximation and (iii) unquenching---they have dynamical fermion
results but do not yet perform a continuum extrapolation with
them. Their results suggest that unquenching will raise the value of
the decay constants. This agrees with an
estimate,\cite{sharpe-zhang,sharpe:lat96} using the difference between
chiral loop contributions in quenched and unquenched QCD, that
$f_{B_s}$ in full QCD is increased by 20\% over its quenched
value. Calculations of $f_B$ in the static limit extrapolated from
negative numbers of flavours~\cite{bermions3} also suggest an increase
of about 20\%.

The SGO collaboration~\cite{sgo:fB-dynam-W,sgo:fB-dynam-SW-TI} is
calculating $f_B$ using a lattice NRQCD action (and heavy-light axial
current) corrected to $O(1/M)$ where $M$ is the heavy quark
mass. However, the renormalisation constants required to match onto
full QCD are not yet available, so I will not quote results for the
decay constant.

The last row of table~\ref{tab:pscalar-decay-consts}
summarises~\cite{jmf-hlwme:lat96} global results for $B$ and $D$ meson
decay constants, following the compilation by
G.~Martinelli.~\cite{marti:heavy-flavours,marti:beauty96,marti:ichep96}

\section{$B$--$\bar B$ Mixing Parameter}

Recent lattice calculations of the $B$ meson mixing parameter
$B_B(\mu)$, defined analogously to the kaon mixing parameter $B_K$ in
Eq.~\ref{eq:BKdefn} below, have been made both for relativistic
quarks~\cite{soni:lat95,bbs:lat96} and in the static
limit.\cite{ukqcd:staticB,ape:BBstatic,ken:BBstatic} The latest static
results~\cite{ape:BBstatic,ken:BBstatic} use new calculations of the
full-theory/static-theory matching incorporating previously omitted
contributions.\cite{cfg:nlo,buchalla:nlo} Since $B_B(\mu)$ is
scale-dependent it is conventional to quote a renormalisation group
invariant (RGI) quantity, $\hat B_B$. At one loop $\hat B_B =
\alphas(\mu)^{-2/\beta_0} B_B(\mu)$, but the two-loop relation is
commonly used.\cite{bjw,buras:ichep96}

Calculations with relativistic quarks show no obvious lattice spacing
dependence and cluster around a value of $\hat B_B = 1.3(1)$ [with a
weighted average of $1.31(3)$, quoted in Warsaw] for the one-loop $B$
parameter,\cite{jmf-hlwme:lat96,ape:BBstatic} or $\hat B_B = 1.4(1)$
using the two-loop formula.\cite{ape:BBstatic} Allowing for
uncertainty in the continuum extrapolation, I quote the two-loop $\hat
B_B = 1.4(1)$ as the quenched lattice result.

The static results differ for calculations using
Wilson~\cite{ken:BBstatic} or SW~\cite{ukqcd:staticB,ape:BBstatic}
light quarks. The Kentucky group~\cite{ken:BBstatic} (Wilson) find
$\hat B_B = 1.40(6)$ for the one-loop RGI quantity, rising to
$1.50(6)$ for the two-loop case.  The APE two-loop
result~\cite{ape:BBstatic} (SW) is $\hat B_B = 1.08(6)(8)$. The
dominant uncertainty is from higher order terms in the perturbative
matching to continuum QCD. Nonperturbative renormalisation will be
crucial to reduce systematic errors.\cite{ape:BBstatic,gmst}

The relevant quantity for $B$--$\bar B$ mixing is $\xi_B^2 = f_B^2
\hat B_B$. Taking the two-loop relativistic quark result for $\hat
B_B$ with $f_B = 175(25)\mev$ from
table~\ref{tab:pscalar-decay-consts} gives $\xi_B = 207(30)\mev$ as
the current lattice estimate.\cite{marti:beauty96} This quantity can
also be extracted directly from the $\Delta B{=}2$ matrix element, $M
= 8 f_B^2 B_B m_B^2 /3$. To avoid uncertainties from setting the
scale, it is convenient to determine the ratio~\cite{bbs:lat96}
\[
r_{sd} = {f_{B_s}^2 B_{B_s} m_{B_s}^2 \over
 f_B^2 B_B m_B^2}
\]
For relativistic quarks a direct extraction gives $r_{sd} =
1.54(13)(32)$. Previously, $r_{sd}$ has been evaluated from separate
results for the decay constant and $B$ parameter ratios. Combining
$f_{B_s}/f_B = 1.15(5)$ from table~\ref{tab:pscalar-decay-consts} with
$B_{B_s}/B_B = 1.01(4)$~\cite{soni:lat95} and $m_{B_s}/m_B = 1.017$
gives $r_{sd} = 1.38(13)$. In the static case, APE~\cite{ape:BBstatic}
find $r_{sd} = 1.35(5)$ by the direct method, or $r_{sd} = 1.43(7)$
from combining $f_{B_s}/f_B = 1.17(3)$ and $B_{B_s}/B_B = 1.01(1)$
measured on the same gauge configurations. The results of the two
methods are quite consistent, but future calculations should improve
on the precision of $r_{sd}$ obtained directly.

The above results are from quenched calculations. Unquenching is
expected to increase $f_{B_s}/f_B$ by about 10\% ($0.16$ from a chiral
loop estimate~\cite{sharpe-zhang,sharpe:lat96}). For $B_{B_s}/B_B$,
numerical evidence suggests a small increase on two-flavour dynamical
configurations~\cite{soni:lat95} but the chiral loop
estimate~\cite{sharpe-zhang,sharpe:lat96} is for a decrease of $-0.04$
in the ratio.

\section{Kaon $B$-Parameter $B_K$}

The parameter $B_K$ is defined by
\be\label{eq:BKdefn}
B_K(\mu) = {\langle\bar K^0 | \bar s \gamma_\mu L d 
  \bar s \gamma^\mu L d | K^0 \rangle \over
  8\langle\bar K^0| \bar s \gamma_\mu L d \ket0
   \langle 0| \bar s \gamma^\mu L d \ket K \!/ 3},
\ee
where $L = 1{-}\gamma_5$. It is a scale dependent quantity for which
lattice results are most often quoted after translation to the value
in $\overline{\rm MS}$ using naive dimensional regularisation (NDR) at
a scale $\mu=2\gev$. I will follow this practice while discussing the
lattice results and convert at the end to the renormalisation group
invariant parameter $\bkhat$ normally used in phenomenology. At
next-to-leading order, $\bkhat$ is given by,
\[
{\bkhat\over B_K(\mu)} = \alphas(\mu)^{{-\gamma_0\over2\beta_0}}
 \Big( 1 + {\alphas(\mu)\over4\pi}{(\beta_1\gamma_0-\beta_0\gamma_1)
 \over 2\beta_0^2} \Big),
\]
where $\beta_{0,1}$ and $\gamma_{0,1}$ are the first two coefficients
of the beta function and anomalous dimension,
respectively.\cite{bjw,cfmr}

Systematic errors in $B_K$ calculations are being carefully
explored. Calculations using staggered fermions (an alternative
formulation of relativistic lattice fermions) are statistically more
precise, but Wilson fermion results are rapidly improving. Here I
summarise the current situation. For more details see the report by
S.~Sharpe from Lattice 96.\cite{sharpe:lat96}

\subsection{Staggered $B_K$}

\begin{figure}
\hbox to\hsize{\hss
\psfig{figure=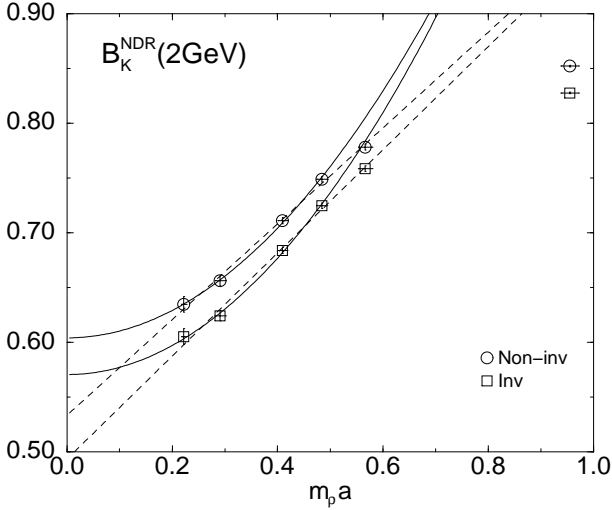,width=\hsize}\hss}
\caption[]{JLQCD results~\protect\cite{jlqcd-BK-stag:lat96} for 
$\bktwo$ as a function of $m_\rho a$, with a quadratic fit to four
leftmost data points (solid) and a linear fit to five leftmost data
points (dashed). ``Non-inv'' and ``Inv'' refer to two possible
discretisations of the lattice operators.}
\label{fig:jlqcd-BK-stag}
\end{figure}

Discretisation errors for $B_K$ using staggered fermions are known to
be $O(a^2)$.\cite{sharpe-BK-stag:lat93} The first
calculation~\cite{sharpe-BK-stag:lat93} performed with a range of
lattice spacings $a$ therefore used a quadratic extrapolation in $a$
and found $\bktwo = 0.616(20)(27)$ for the quenched result.  The data
itself, however, could not distinguish linear and quadratic $a$
dependence. New results from the JLQCD
collaboration~\cite{jlqcd-BK-stag:lat95,jlqcd-BK-stag:lat96} are shown
in Fig.~\ref{fig:jlqcd-BK-stag}.  Their data fits better to a linear
than a quadratic dependence on $a$. The argument for $O(a^2)$
corrections has been
checked,\cite{jlqcd-BK-stag:lat96,luo:improvement,luo:lat96} however,
so results are quoted for a quadratic fit.\cite{jlqcd-BK-stag:lat96}
Future calculations at smaller $a$ should confirm the leading $a^2$
dependence. A continuum-extrapolated quenched result has also been
given by a group from OSU.\cite{osu:BK-unquenched} The new results
are ($n_f$ is the number of dynamical flavours):
\be\label{eq:BKstag}
%\bktwo^{n_f=0}_{a\to0} =
\bktwo_{{a\to0\atop n_f=0}} =
\cases{0.587(7)(17)&JLQCD\cr 0.573(15)&OSU\cr}
\ee
I will take these as the best quenched lattice estimates of $B_K$.

Two issues need addressing to relate the results in
Eq.~(\ref{eq:BKstag}) to $B_K$ for full QCD. One is the inclusion of
dynamical quarks. The second is to allow $m_s\neq m_d$, since the
calculations above have a kaon composed from degenerate quarks.

Important progress in unquenching $B_K$ has been made by the OSU
group,\cite{osu:BK-unquenched} who find a statistically significant
increase in $B_K$ (earlier studies suggested a small
decrease~\cite{soni:lat95}),
\be\label{eq:BKunquench}
{\bktwo^{n_f=3}\over\bktwo^{n_f=0}} = 1.05(2).
\ee
They calculated with $n_f = 0,2,4$ for fixed lattice spacing
$a=2\gev$. Sharpe~\cite{sharpe:lat96} gives a more conservative
estimate for the ratio as $1.05(15)$ after allowing for the
extrapolation to $a{=}0$.

To calculate $B_K$ with physical mass non-degenerate quarks requires
fermions with very small masses and will therefore be
difficult. However, chiral perturbation theory fixes the quark mass
dependence of $B_K$, providing a way to determine the non-degeneracy
correction.\cite{bsw:BK-chpt,sharpe:prd46} The calculation has yet to
be done, so Sharpe~\cite{sharpe:lat96} estimates,
\be\label{eq:BKnondegen}
B_K^{\rm nondegen} = 1.05(5)\, B_K^{\rm degen}.
\ee

Combining the staggered fermion results in Eq.~(\ref{eq:BKstag}) with
the unquenching and nondegeneracy corrections in
Eqs.~(\ref{eq:BKunquench}) and~(\ref{eq:BKnondegen}) leads to the final
estimate: \be\label{eq:BKfinal} \bktwo = 0.64(2)(10).  \ee The first
error is that in the quenched value. The second is the larger 15\%
unquenching error combined in quadrature with the 5\% error for
nondegeneracy. Since Eq.~(\ref{eq:BKfinal}) incorporates estimates of
systematic effects in the central value, an alternative statement is
$\bktwo = 0.58(2)(9)$, where the central value is the quenched result,
noting that unquenching and nondegeneracy can raise the value by
10\%. Converting the result in Eq.~(\ref{eq:BKfinal}) to $\bkhat$
using $\alphas(2\gev)=0.3$ and three flavours gives~\footnote{This
differs from $0.90(6)$ quoted in Warsaw. It uses updated
JLQCD~\protect\cite{jlqcd-BK-stag:lat96} and new
OSU~\protect\cite{osu:BK-unquenched} results with a more
conservative~\protect\cite{sharpe:lat96} unquenching and degeneracy
error.}
\[
\bkhat = 0.87(3)(14).
\]

\subsection{Wilson $B_K$}

Calculations of $B_K$ using Wilson and SW fermions have to deal with
the explicit breaking of chiral symmetry by the fermion action. This
means that the continuum operator of interest mixes with four other
dimension six lattice operators:
\[
{\cal O}^{\rm cont} = Z \big( {\cal O}^{\rm latt} +
 \sum_{i=1}^4 z_i {\cal O}_i \big) + O(a).
\]
The constants $Z$ and $z_i$ have to be determined so that lattice and
continuum matrix elements agree to $O(a)$. A general four-fermion
operator has the chiral expansion
\begin{eqnarray*}
\lefteqn{\langle \bar K | {\cal O} \ket K =
    \alpha + \beta m_K^2 + \delta_1 m_K^4 +}& \\
 & & p_{\bar K}\mathord\cdot p_K(\gamma + \delta_2 m_K^2 +
                  \delta_3 p_{\bar K}\mathord\cdot p_K) + \cdots.
\end{eqnarray*}
For the continuum operator used to determine $B_K$, chiral symmetry
demands that $\alpha = \beta = \delta_1 = 0$. The vanishing of these
momentum-independent lattice artifacts can be used to test that the
$z_i$ are correct.

Various methods have been applied to determine the $z_i$ and $Z$.  One
loop perturbation theory gives incorrect chiral behaviour. One can
adjust the $z_i$ by hand to restore the correct
behaviour~\cite{bs-BK:lat89,soni:lat95} or calculate with varying
momenta to isolate and discard the
artifacts.\cite{lanl:wme-lat95,gb-mq:lat96} A better procedure is to
calculate $Z$ and the $z_i$ nonperturbatively. The Rome group demand
that quark matrix elements satisfy continuum normalisation
conditions,\cite{mpstv,dmstv,rossi:lat96} while
JLQCD~\cite{jlqcd-BK-WID:lat96} impose chiral Ward identities on quark
matrix elements to determine the $z_i$, together with a continuum
normalisation step to fix $Z$. Both groups find that the matrix
element of ${\cal O}^{\rm cont}$ is obtained with the correct chiral
behaviour. The errors are currently larger than for staggered
fermions, principally because of the need to calculate several
constants $z_i$.

JLQCD~\cite{jlqcd-BK-WID:lat96} find that the continuum extrapolation
is best done for a quantity differing from $B_K$ by lattice artifacts
which vanish as $a{\to}0$. Extrapolating from results at three
different $a$ values they find $\bktwo^{n_f=0}_{a\to0} = 0.59(8)$, in
excellent agreement with the staggered results in
Eq.~(\ref{eq:BKstag}).

\section{Strong Coupling $\alphas$}

Determinations of $\alphas$ from Lattice QCD are done in three steps:
(i) define and measure some $\alphaslatt$, (ii) determine the lattice
spacing $a$, to set the scale at which $\alphaslatt$ takes its
measured value, and (iii) convert to $\alphasmsbar$. Step (i) is
necessary because the bare lattice coupling, determined from the
simulation parameter $\beta = 6/g^2$, is rather small and has a
badly-behaved perturbation theory. One must instead use a physical
definition for $\alphaslatt$. Steps (ii) and (iii) are the major
source of uncertainty. Here I will give an update and refer the reader
to recent reviews for more
details.\cite{cmi:lepton-photon,shige:lat96,elkhadra-moriond96}  New
results are available~\cite{shige:lat96} from the NRQCD collaboration
and a Fermilab-SCRI group, both using quarkonium level splittings to
fix the lattice spacing. Other lattice methods of determining
$\alphas$ are being developed. These include studying the three gluon
vertex~\cite{threegluon} and a program by the ALPHA collaboration
using the Schr\"odinger Functional method, outlined by S.~Sint at
this conference.\cite{pa15-010}

\subsection{$\alphas$ from Quarkonia}

The measured lattice coupling
is $\alphaP$, defined exactly by~\cite{nrqcd-alpha:lat95}
\[
-\ln W_{1,1} = {4\pi\over 3} \alphaP\big({3.4\over a}\big)
 \big( 1 - [1.19+0.07 n_f]\alphaP \big).
\]
$W_{1,1}$ is the single plaquette expectation value which can be
determined accurately for $n_f{=}0$ and $2$ with varying sea quark
masses.

The lattice spacing is determined from 1S--1P and 1S--2S quarkonium
level splittings which are known experimentally to be very insensitive
to the heavy quark mass in the bottom to charm region. Quarkonia are
tiny systems, so finite volume effects should be small. The actions
used can be systematically improved to control discretisation
errors. Unquenching effects can be estimated by extrapolating
$\alphaP^{-1}$ linearly in $n_f$. A further systematic error is the
effect of the sea quark mass on the level splittings. Sea quarks in
lattice simulations are heavier than physical up or down quarks.
Grinstein and Rothstein~\cite{grinroth} estimate that the
extrapolation to physical masses could increase the final value for
$\alphasmsbar(m_Z)$ by $0.003$.

To convert to a continuum coupling, one uses:
\[
\alphasmsbar^{(n_f)}(Q) = \alphaP^{(n_f)}(Qe^{5/6})
 (1 + {2\over\pi}\alphaP + C_2^{(n_f)}\alphaP^2 + \cdots).
\]
The $n_f$-dependent constant $C_2$ is now known~\cite{lw:C2} in the
quenched case, $C_2^{(n_f=0)} = 0.96$. It was previously set to
zero. Using the quenched value even for $n_f=3$ raises the result for
$\alphasmsbar(m_Z)$, although there is clearly still a systematic
error here.

\begin{figure}
\hbox to\hsize{\hss
\psfig{figure=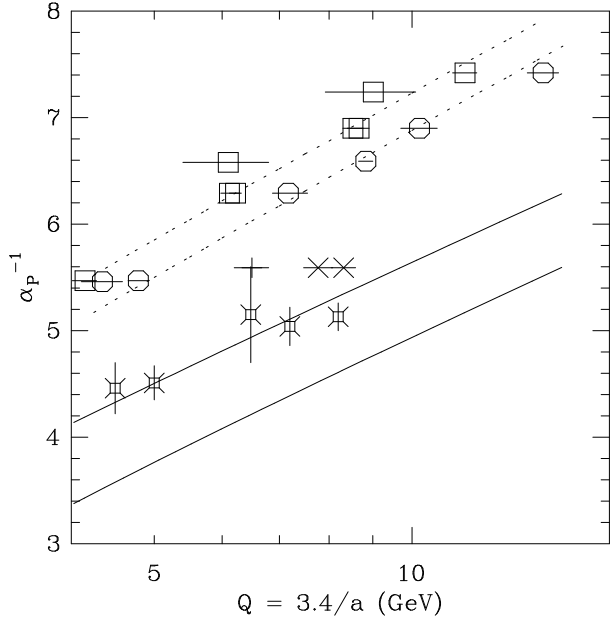,width=\hsize}\hss}
\caption[]{Recent lattice determinations of $1/\alphaP$ as a function of
the scale $Q=3.4/a$ at which they are
measured~\protect\cite{shige:lat96}. Boxes denote quenched ($n_f{=}0$)
results with the scale set from charmonium and circles are quenched
results with the scale from $\Upsilon$. Two-flavour ($n_f{=}2$) results
from charmonium and $\Upsilon$ are denoted by pluses and crosses
respectively. Results extrapolated to $n_f{=}3$ are given by fancy
boxes. The upper and lower solid curves correspond to
$\alphasmsbar(m_Z) = 0.115$ and $0.125$ respectively.}
\label{fig:alphaP}
\end{figure}
Fig.~\ref{fig:alphaP} shows recent measurements of $\alphaP$,
plotted as $\alphaP^{-1}$ against the value $Q=3.41/a$ at which they
are measured.\cite{shige:lat96} The solid curves superimposed on the
figure show what value for $\alphaP^{-1}(Q)$ for $n_f=3$ would convert
to given values for $\alphasmsbar(m_Z)$. The most recent results
from the NRQCD collaboration give~\cite{shige:lat96}:
\[
\alphasmsbar(m_Z) = \cases{0.1175(11)(13)(19)&1S--1P,\cr
 0.1180(14)(14)(19)&1S--2S.\cr}
\]
The first error is a combination of statistics, determination of the
lattice spacing $a$ and relativistic corrections. The second error
comes from the extrapolation in the sea quark mass and the third,
dominant, error is from the conversion to $\alphasmsbar$ (from the
difference between using $C_2=0$ and $0.96$). These combine to
give $ \alphasmsbar(m_Z) = 0.118(3)$.  The NRQCD method currently
relies on perturbation theory to fix the coefficients in the action,
for which it is hard to estimate systematic errors. Nonperturbative
renormalisation techniques may help here. The Fermilab/SCRI group use
a different effective action to calculate the $\bar b b$ and $\bar c
c$ spectra in the quenched approximation and the $\bar b b$ spectrum
for $n_f{=}2$. Applying the same analysis as above yields a
preliminary result~\cite{shige:lat96} of $\alphasmsbar(m_Z) =
0.116(3)$. Combining the NRQCD and Fermilab/SCRI results I quote
\[
\alphasmsbar(m_Z) = 0.117(3).
\]

\section{Light Quark Masses}

The masses of the light quarks, $m_u$, $m_d$ and $m_s$, are three of
the least well known standard model parameters, but their values are
important in a number of areas.  The strange quark mass, for example,
appears in the evaluation of matrix elements for the $\Delta I=1/2$
rule and the CP-violation parameter $\epsilon'/\epsilon$. Chiral
perturbation theory allows the extraction of the mass ratios from
pseudoscalar meson masses. QCD sum rules can be applied for the masses
themselves, but rely on detailed experimental information about the
hadronic spectral function. Direct calculation of the masses from
lattice QCD is thus an important challenge.

Lattice simulations determine a bare lattice-spacing-dependent quark
mass $m(a)$ which can be related to a continuum renormalised mass,
such as the $\overline{\rm MS}$ mass, $\mmsbar(\mu)$. The conversion
factor can be calculated using (boosted) perturbation theory. It has
become standard to quote results for $\mmsbar(2\gev)$, using a scale
which matches the typical scale of lattice calculations ($a^{-1} \sim
2$--$4\gev$) and for which perturbation theory should work better.
The lattice mass is determined by evaluating pseudoscalar or vector
meson masses, for which the mass-squared or mass itself depend
linearly on the quark mass respectively.

\begin{figure}
\hbox to\hsize{\hss
\psfig{file=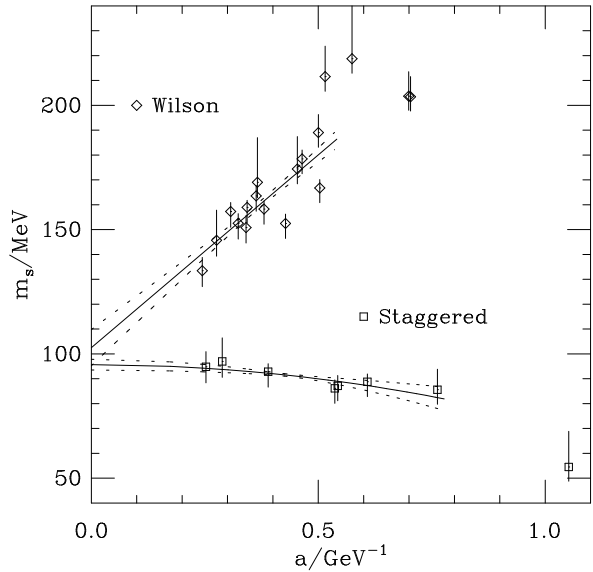,width=\hsize,%
bbllx=25pt,bblly=37pt,bburx=516pt,bbury=505pt}
\hss}
\caption[]{Strange quark mass, $\mmsbar_s(2\gev)$, extracted using
$m_\phi$ in quenched calculations with Wilson and staggered
quarks~\protect.\cite{gb-lightquarks,rajan} Also shown are fits for
the continuum extrapolation of both sets of points.}
\label{fig:ms}
\end{figure}
Gupta and Bhattacharya (G\&B)~\cite{gb-lightquarks,rajan} have made a
recent global analysis, performing a continuum extrapolation, $a\to0$,
of quark masses calculated using both Wilson and staggered
fermions. The extrapolation for the strange quark mass is shown in
Fig.~\ref{fig:ms}. For Wilson quarks the leading lattice spacing
dependence is $O(a)$ and for staggered quarks it is $O(a^2)$. A
Chicago-Fermilab-Hiroshima-Illinois (CFHI)
group~\cite{fnal-mq:lat96,fnal-lightquarks} compare results for Wilson
quarks and quarks with a tadpole-improved SW action, where $O(a)$
effects are reduced, first appearing as $O(\alphas a)$, and also
perform a continuum extrapolation.

Results for the strange quark mass from quenched calculations are:
\[
\mmsbar_s(2\gev) =
\cases{90(20)(10)\mev&G\&B~\cite{gb-lightquarks,rajan}\cr
       95(16)\mev&CFHI~\cite{fnal-lightquarks}\cr
      %128(18)\mev&APE~\cite{ape-quarkmass}\cr
      %141(17)\mev&APE NPT~\cite{ape-quarkmass}\cr
      }
\]

For the $u$ and $d$ quarks, the quantity extracted is the average
mass, $\bar m = (m_u{+}m_d)/2$. Quenched results for this are:
\[
\bar m^{\overline{\rm MS}}(2\gev) =
\cases{3.2(4)(3)\mev&G\&B~\cite{gb-lightquarks,rajan}\cr
       3.6(6)\mev&CFHI~\cite{fnal-lightquarks}\cr
      }
\]

These values are already at the bottom end of the range predicted by
other methods. Since $m_s$ appears quadratically in the evaluation of
matrix elements for $\epsilon'/\epsilon$, a low value could have
important implications for standard model calculations of CP
violation.\cite{buras:ichep96}

The Wilson data for the strange quark mass clearly depend on the
lattice spacing. Although the leading $a$ dependence is linear, it
could be that a number of effects are conspiring to produce an
apparent linear dependence in the current data (compare the case for
$B_K$ for staggered fermions, where the leading $O(a^2)$ is satisfied,
if at all, only for the data at the smallest lattice spacings used to
date). In this case, the final result may well turn out to be
different.  These values rely on a perturbative matching between
lattice and continuum definitions of the quark masses, which could
prove to be unreliable although the perturbative correction is not
large. Nonperturbative methods have been proposed and tested, based on
the Ward identity for the axial vector current.\cite{ape-quarkmass}

The results above are all in the quenched approximation. Some
calculations are also available with two flavours of dynamical
fermions. For the same lattice spacing they lie systematically below
the quenched results, an effect anticipated from the different running
of the strong coupling in quenched and unquenched
QCD.\cite{mackenzie:lat93} The reduction of the quark masses in full
QCD might be very dramatic, but I believe the current paucity of data
forbids meaningful numerical predictions. It will be extremely
interesting, however, to follow developments in these calculations.

\section{The Lightest Scalar Glueball}

Experiments~\cite{landua:spectroscopy} show that there are more
scalar-isoscalar resonances with masses below $1.8\gev$ than can be
accounted for by light $q\bar q$ states. This is strong evidence that
glueballs exist and two prime candidates are the $f_0(1500)$ and
$f_0(1720)$ mesons.\cite{fec:glueball-picture} Since glueball and
$q\bar q$ states are expected to mix, the glueball content of these
candidates remains to be determined.

Lattice calculations provide input by calculating glueball masses in
the quenched approximation, where glueballs are stable and do not mix
with quark states. In principle, full QCD can be modelled on the
lattice, tuning the sea quark masses between physical values, where
the experimental meson spectrum should be reproduced, and large
values, where one matches onto quenched results. In practice, glueball
studies using dynamical quarks, where glueball-meson mixing can occur,
are just beginning. Initial results~\cite{sesam-glue:lat96} suggest
that the masses will not change by more than about 10\%. The dynamical
masses used are still quite large, however, and things will become
more complicated when these masses are light enough to allow glueball
decay.

Two groups have continuum-extrapolated quenched results for the
lightest scalar glueball:
\[
m_{0^{++}} = \cases{1550(50)\mev&UKQCD~\cite{ukqcdballs}\cr
            1740(71)\mev&GF11~\cite{gf11-glueball:lat93}\cr}
\]
Continuum extrapolations of all available data can differ because
different quantities are used to fix the lattice spacing and different
ranges of $a$ may be used. Results are:
\[
m_{0^{++}} = \cases{1600(160)\mev&
                 Michael~\cite{cmi:lepton-photon,cmi:lathuile}\cr
            1707(64)\mev&Weingarten~\cite{weingarten:lat96}\cr
            1610(150)\mev&
                 Close \& Teper~\cite{fec:glueball-picture,pa01-113}\cr}
\]
Quenched lattice calculations also now exist for scalar $s\bar s$
quarkonium.\cite{ukqcd:orb-hyb,lee-wein:lat96} These have not been
extrapolated to zero lattice spacing, but the evidence is that the
quenched scalar $s\bar s$ mass is below the quenched scalar glueball
mass.

The lattice results can be used as input for glueball-quarkonium
mixing models. Weingarten~\cite{weingarten:lat96} has a simple model
mixing the glueball with $s\bar s$ and $n\bar n \equiv (u\bar u +
d\bar d)/\sqrt2$. The observed $f_0(1390)$, $f_0(1500)$ and
$f_0(1720)$ masses (if the latter state is confirmed as a
scalar) can be reproduced for input glueball and $s\bar s$ masses of
about $1640\mev$ and $1520\mev$ respectively, consistent with the
lattice results. The $f_0(1720)$ is more than 75\% glueball (in
probability), while the $f_0(1500)$ is more than 75\% $s\bar
s$. Moreover, the $n\bar n$ component of the $f_0(1500)$ has opposite
sign to the $s\bar s$ component, which could help explain the observed
suppression in the width of $f_0(1500)$ decays to $K\bar K$.

A quenched calculation has also been made of the coupling of the
$0^{++}$ glueball to two pseudoscalar mesons.\cite{gf11-couplings}
This is a very delicate calculation made at a single value of the
lattice spacing. The result, however, is a value of $108(29)\mev$ for
the total two-body width, implying that the lightest scalar glueball
should be easy to find experimentally. The calculation also indicates
that the two-body couplings increase with increasing pseudoscalar
mass, consistent with observations for the $f_0(1720)$.

The Weingarten model and arguments suggest that the $f_0(1720)$ is
predominantly a glueball and $f_0(1500)$ is predominantly $s\bar s$
quarkonium. However, other models produce different
conclusions.\cite{close-amsler,fec:glueball-picture} Various
experimental tests to determine the flavour content of the isoscalar
mesons have been
proposed.\cite{pa01-113,close-amsler,farrar:ichep96,cgl} Future
lattice calculations, models and experiments should help pin down the
glueball.

\section*{Acknowledgments}
I thank the conference organisers for their invitation to speak, the
Royal Society for a grant towards attendance costs and the following
for assistance in various ways: C.~Allton, G.~Bali, A.~Buras,
F.~Close, R.~Gupta, A.~El~Khadra, A.~Kronfeld, L.~Lellouch, G
Martinelli, C.~Michael, C.~Sachrajda, S.~Sharpe, J.~Shigemitsu,
J.~Simone, A.~Soni, D.~Weingarten, M.~Wingate and H.~Wittig. Where
results have been updated since the conference, I have endeavoured to
include them in this review. Work supported by the Particle Physics
and Astronomy Research Council under grant GR/K55738.

\section*{References}
\bibliographystyle{ichep}
\bibliography{lat96,ichep96,pa,lat93}

\end{document}